%% file: FTSMC_arXiv_v2.tex
\documentclass[journal]{IEEEtran}
\usepackage{amsmath}
\usepackage{amssymb}
\usepackage{amsfonts}
\usepackage{cite}
\usepackage{graphicx}
\usepackage{subfig}
\usepackage{float}
\usepackage{color}
\usepackage{cite}
\include{BoldSymbol}  

\newcommand{\FB}{\mbox{$\mathcal{F}_B$}}
\newcommand{\FI}{\mbox{$\mathcal{F}_I$}}

\newcommand{\hR}{\mbox{$\mathbb{R}$}}

\newcommand{\hS}{\mbox{$\mathbb{S}$}}

\newcommand{\hZ}{\mbox{$\mathbb{Z}$}}
\newcommand{\vone}{\mbox{$\mathbf{1}$}}

\newtheorem{theorem}{Theorem}[section]
\newtheorem{assumption}{Assumption}[section]
\newtheorem{lemma}{Lemma}[section]

\newtheorem{remark}{Remark}[section]
\newtheorem{algorithm}{Algorithm}[section]

\begin{document}
\graphicspath{ {images_20190430/} }
\title{Observer-Based Fault-Tolerant Spacecraft Attitude Tracking Using Sequential Lyapunov Analyses}

\author{Haichao~Gui\thanks{Haichao~Gui, Associate Professor, School of Astronautics, Beihang University, Beijing, P.~R. China (email: hcgui@buaa.edu.cn)}
        \thanks{The work in this paper has received funding from the National Natural Science Foundation of China under grants 11702010 and 11972056, and from the Research Fund of the Science and Technology on Space Intelligent Control Laboratory under grant KGJZDSYS-2018-06}}


\maketitle

\begin{abstract}
The spacecraft attitude tracking problem is addressed with actuator faults and uncertainties among inertias, external disturbances, and, in particular, state estimates. A continuous sliding mode attitude controller is designed using attitude and angular velocity estimates from an arbitrary stable stand-alone observer. Rigorous analysis shows that the controller ensures robust stability of the entire closed-loop system as long as the observer yields state estimates with uniformly ultimately bounded estimation errors. In addition, a sequential Lyapunov analysis is utilized to obtain a convergent sequence of analytical, successively tighter upper bounds on the steady-state tracking error. Therefore, our results can be used to predict steady-state performance bounds given selected gains or facilitate gain selection given steady-state performance bounds. Numerical examples demonstrate the utility of the proposed theory.
\end{abstract}


%
\IEEEpeerreviewmaketitle

\section{Introduction}

The attitude control of a rigid spacecraft has attracted extensive attention. The attitude estimation and control problems are coupled because the attitude and angular velocity estimates required for implementing a certain feedback controller are generated by some observers/filters. In spite of this, they are treated separately at most times to avoid difficulties in theoretical analysis and simplify designs. As such, substantial feedback control laws have been designed using various methods in the past decades by assuming that perfect knowledge about the attitude and angular velocity is known and a separation principle holds between the controller and observer (see, e.g.,  \cite{Wen:91, Sanyal:09, Mayhew:11, Lee:15TAC, Gui:17RNC2}). Both assumptions, however, are not true. Attitude sensors always contain measurement errors that will lead to imprecise state estimates. In addition, the attitude equations are nonlinear and there exists no general separation principle for nonlinear systems.

Observer-based attitude control has been investigated in \cite{Thienel:03, Akella:15, Gui:16SCL, Wu:16, Berkane:18TAC1} with established closed-loop stability. The corresponding stability results, however, are limited to the specific observer and controller considered therein. Additionally, most of these studies assumed the absence of any system uncertainties, except for \cite{Thienel:03}, which took measurement errors into account. Recently, de Ruiter \cite{deRuiter:13, deRuiter:16TAC} has presented a backstepping controller and an adaptive controller that can be combined with a series of stable stand-alone observers for attitude tracking while bearing robustness against uncertain inertias, disturbance torques, and measurement errors. In \cite{Gui:18JGCD2}, the classical quaternion proportional-derivative attitude controller was shown to possess similar properties. The studies in \cite{Thienel:03, deRuiter:13, deRuiter:16TAC, Gui:18JGCD2}, however, all ignored the uncertainty due to actuator faults \cite{Hamayun:12, Shen:15AUTO, HuQL:19,Liu:18}, which can commonly occur among real applications.

This paper investigates the observer-based attitude control of a fully-actuated rigid body with uncertain inertias, external disturbances, uncertain state estimates, and actuator faults. Similarly to \cite{deRuiter:16TAC}, no specific observer is considered. Instead, it is assumed that the attitude and angular velocity estimates are generated from any standalone observers that ensure uniformly ultimately bounded (UUB) estimation errors. We propose a continuous sliding mode controller that achieves attitude tracking with proven robustness against all the aforementioned system uncertainties. Moreover, by means of a sequential Lyapunov technique we derived a convergent sequence of analytical, successively tighter upper bounds on the ultimate state tracking errors. Following this, a numerical algorithm is developed to compute less conservative bounds on the ultimate the state tracking errors, provided that some upper bounds on the system uncertainties are available. The results can be used to predict steady-state performance bounds for selected gains or facilitate gain selection given steady-state performance bounds. Numerical examples are presented to validate our analyses and demonstrate the utility of the proposed algorithm.

The rest of this paper is organized as follows. The spacecraft attitude equations and fault-tolerant attitude control problem are formulated in Section II. Section III derives an observer-based sliding mode controller with rigorous stability proof and an algorithm to estimate the ultimate attitude tracking accuracy. The application of the proposed method is illustrated via numerical examples in Section IV and some remarks are concluded in Section V.

\textit{Notations:} Throughout the paper, denote by $\|\cdot\|$ the 2-norm of a vector or matrix, and $\lambda_{\textup{min}}(\cdot)$ and $\lambda_{\textup{max}}(\cdot)$ the minimum and maximum eigenvalues of a real symmetric matrix, respectively. Denote by $\hS^3 = \{\bcq =[q_0, \bsq^T]^T\in\hR^4: q_0^2 + \bsq^T \bsq =1 \}$ the set of unit quaternions. The multiplication between $\bcq =[q_0, \bsq^T]^T, \bcp=[p_0,\bsp^T]^T \in \hS^3$ is defined as \cite{Sidi:00}

\begin{equation*}
\bcq\otimes \bcp=
\begin{bmatrix}
q_0{p_0} - \bsq^{T}\bsp \\
q_0 \bsp + p_0\bsq + \bsq\times \bsp
\end{bmatrix}
\end{equation*}
In the following, the time argument is omitted when there is no confusion.

\section{Problem Formulation}

\subsection{Equations of Attitude Motion}

The attitude kinematics and dynamics of a rigid spacecraft with respect to the inertial frame {\FI} can be written as

\begin{equation}\label{eq:quatkine}
\dot{\bcq} = \frac{1}{2}\left[\begin{array}{c}
                     - \bsq^T \\
                     G(\bcq)
                   \end{array}
\right] \bomega, \quad G(\bcq)= q_0 I_3 + \bsq^\times
\end{equation}

\begin{equation}\label{eq:eulerdyn}
J\dot{\bomega} = -\bomega \times J \bomega + \btau_c + \btau_d
\end{equation}
where $\bcq = [q_0,\bsq^T]^T= [q_0, q_1, q_2, q_3]^T\in\hS^3$ describes the attitude of the spacecraft body frame {\FB} relative to the inertial frame {\FI}, $I_n$ is the $n\times n$ identity matrix, $\bomega \in\hR^3$ is the spacecraft angular velocity expressed in {\FB}, $J = J^T\in\hR^{3\times 3}$ is the spacecraft inertia matrix, $\btau_c \in\hR^3$ is the control torque, and $\btau_d \in\hR^3$ is the external disturbance torque. In addition, $\bsq^\times$ is a skew-symmetric matrix given by
\begin{equation*}
\bsq^{\times}=
\left[\begin{array}{ccc}
       0 & -q_3 & q_2 \\
       q_3 & 0 & -q_1 \\
       -q_2 & q_1 & 0
\end{array}\right]
\end{equation*}

The desired attitude and angular velocity are usually generated from some motion planning algorithms and denoted as $\bcq_d\in\hS^3$ and $\bomega_d\in\hR^3$, which also obey the kinematics given in (\ref{eq:quatkine}). Note that for any unit quaternion $\bcq = [q_0, \bsq^T]^T\in\hS^3$, its inverse is given by $\bcq^{-1} = [q_0, -\bsq^T]^T\in\hS^3$. The attitude and angular velocity tracking errors, $\bcq_e = [q_{e0}, \bsq_e^T]^T\in\hS^3$ and $\bomega_e\in\hR^3$, are then defined as

\begin{equation}\label{eq:errvar}
\begin{array}{c}
\bcq_e = \bcq_d^{-1} \otimes \bcq \\
\bomega_e = \bomega - R(\bcq_e)\bomega_d = \bomega - \bar{\bomega}_d, \quad \bar{\bomega}_d \triangleq R(\bcq_e)\bomega_d
\end{array}
\end{equation}
where $R(\bcq_e)$ is the rotation matrix corresponding to $\bcq_e$ and is computed by

\begin{equation} \label{eq:Rqe}
R(\bcq_e) = I_3 - 2 q_{e0}\bsq_e^{\times} + 2\bsq_e^{\times}\bsq_e^{\times}
\end{equation}

Given a constant $k>0$, define a sliding variable

\begin{equation}\label{eq:s}
\bss = \bomega_e + k\bsq_{e}
\end{equation}
which is an alternate characterization of the tracking error. Differentiating $\bcq_e$ and $\bss$ and invoking~(\ref{eq:quatkine}) and (\ref{eq:eulerdyn}), the tracking error equations in terms of $\bcq_e$ and $\bss$ can be written as

\begin{equation}\label{eq:dot:qe}
\dot{\bcq}_e = \frac{1}{2}\left[\begin{array}{c}
                     - \bsq_e^T \\
                     G(\bcq_e)
                   \end{array}
\right] \bomega_e
\end{equation}

\begin{equation}\label{eq:dot:s1}
\begin{array}{lll}
J\dot{\bss} & = & J\dot{\bomega}_e +  k J \dot{\bsq}_{e} = J\dot{\bomega}_e +  \frac{1}{2}k J G(\bcq_e)  \bomega_e  \\
  & = & \Xi(J, \bomega_e, \bar{\bomega}_d)\bss + \frac{1}{2}k [\bsq_{e}^{\times} J + J \bsq_{e}^{\times}] \bss  \\
  & & + \bpsi - \bpsi_d + \btau_c + \btau_d
\end{array}
\end{equation}

\begin{equation}\label{eq:Xi}
  \Xi(J, \bomega_e, \bar{\bomega}_d) = (J(\bomega_e + \bar{\bomega}_d))^{\times} - \bar{\bomega}_d^{\times}J - J\bar{\bomega}_d^{\times}
\end{equation}

\begin{equation}\label{eq:psi}
\begin{array}{l}
 \bpsi = - \frac{1}{2}k^2 \bsq_e^{\times} J \bsq_e + \frac{1}{2}k G(\bcq_e) J \bomega_e - k \Xi(J, 0, \bar{\bomega}_d)\bsq_{e} \\
  \bpsi_d = \bar{\bomega}_d^{\times}J \bar{\bomega}_d + J R(\bcq_e)\dot{\bomega}_d
\end{array}
\end{equation}
where  $\bomega_e = \bss - k \bsq_e$ is utilized to derive~(\ref{eq:dot:s1}) and obtain the terms given in~(\ref{eq:Xi}) and (\ref{eq:psi}). The details can be found in the Appendix.

\begin{figure*}[!tbp]
\begin{center}
\subfloat[]{\includegraphics[height=4cm]{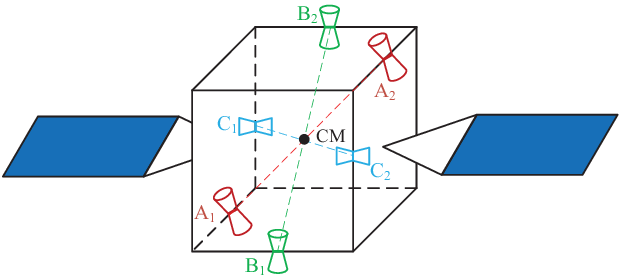}}
\subfloat[]{\includegraphics[height=4cm]{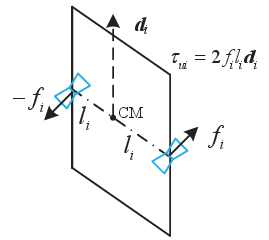}}
\vspace{0 mm}
\caption{Schematic of a spacecraft actuated by reaction thrusters}  \label{fig:SC}
\end{center}
\end{figure*}

\subsection{Observer-Based Fault-Tolerant Control Problem}

Assume that the spacecraft is actuated by multiple reaction thrusters symmetrically distributed about its center-of-mass (CM), as shown in Fig.~\ref{fig:SC}. Thrusters are paired to provide both positive and negative torques along a fixed direction relative to the spacecraft. Figure 1b shows the arrangement of the $i$-th pair of thrusters, which produce reaction forces in opposite directions, i.e., $\pm f_i$. The moment arms of the two forces with respect to the spacecraft CM are both $l_i >0$. Hence, the resultant control torque is $ \tau_{ui}\bsd_i$, where $\tau_{ui} = 2 f_i l_i$ is the torque magnitude and $\bsd_i$ is a unit vector along the output torque direction of the $i$-th pair of thrusters. More knowledge about torques and attitude control by reaction thrusters can be found in \cite{Sidi:00}. Assume that $m \geq 3$ pairs of thrusters are installed so that three control torques can be generated. Let $0 \leq e_i(t)\leq 1$ be the health indicator of the $i$-th pair of thrusters, which are fully functional when $e_i(t)=1$, and are turned off or failed completely if $e_i(t)=0$. When $0 < e_i(t)< 1$, they partially lose their actuating power, which can be caused by faults in thruster valves, fading pressure of the propellent tank, or other reasons. The $i$-th pair of thrusters are supposed to produce control torque $\tau_{ui} \bsd_i$  but the real output torque is $ e_i(t) \tau_{ui} \bsd_i$ due to actuator faults. Hence, $\btau_c$ from actuators can then be written as \cite{Hamayun:12, Shen:15AUTO}

\begin{equation} \label{eq:tauc}
\btau_c = \sum_{i=1}^{m} e_i(t) \tau_{ui} \bsd_i =  D E(t) \btau_u
\end{equation}
where $D = [\bsd_1, \cdots, \bsd_m] \in\hR^{3\times m}$ is the actuator distribution matrix, $E(t) =  \mathrm{diag}(e_1(t), \cdots, e_m(t))\in\hR^{m\times m}$, and $\btau_u = [\tau_{u1}, \cdots, \tau_{um}] \in\hR^m$ is the command torque fed to the actuators.

In order to perform attitude tracking control, the spacecraft attitude $\bcq$ and angular velocity $\bomega$ need to be estimated using certain observers/filters from sensor measurements and the corresponding estimates are denoted as $\hat{\bcq} = [\hat{q}_0, \hat{\bsq}^T]^T \in\hS^3$ and $\hat{\bomega}\in\hR^3$. Our objective is to design the command torque $\btau_u$ using state estimates  $\hat{\bcq}$ and $\hat{\bomega}$ such that the attitude tracking is achieved with guaranteed robustness against state estimation errors, uncertain inertias, unknown external disturbances, and actuator faults. Meanwhile, we intend to provide an efficient algorithm to predict upper bounds on the steady-state attitude tracking errors. The aforementioned system uncertainties are described in the following.

Assume that $\hat{\bcq}$ and $\hat{\bomega}$ are generated by a stable stand-alone observer (e.g., those in \cite{Seo:07, Gui:16SCL, Wu:16, Akella:15, Gui:17RNC1, Thienel:03, Yang:16, Berkane:18TAC2}) and satisfy

\begin{equation} \label{eq:hat:qw}
\hat{\bcq} = \bcq \otimes \tilde{\bcq}^{-1}, \quad  \hat{\bomega} = \bomega + \tilde{\bomega}
\end{equation}
where the estimation errors $\tilde{\bcq} = [\tilde{q}_0, \tilde{\bsq}^T]^T \in\hS^3$ and $\tilde{\bomega}\in\hR^3$ are continuous and bounded as assumed in the following.

\begin{assumption} \label{assum1}
$\tilde{\bsq}(t)$ and $\tilde{\bomega}(t)$ are continuous and uniformly bounded, and satisfy
\begin{equation*}
\limsup_{t\to\infty}\|\tilde{\bsq}(t)\| \leq \rho_q, \quad \limsup_{t\to\infty}\|\tilde{\bomega}(t)\| \leq \rho_w
\end{equation*}
for some constants $0 \leq \rho_q<1$ and $ \rho_w \geq 0$.
\end{assumption}

Since two quaternions $\pm\tilde{\bcq}$ correspond to the same attitude estimation error, it is reasonable to assume that $\tilde{q}_0(t)$ eventually becomes negative \cite{deRuiter:13} and thus
\begin{equation} \label{eq:assum0}
\limsup_{t\to\infty} \tilde{q}_0(t) \geq \sqrt{1 - \rho_q^2}
\end{equation}


Denoted by  $\hat{J} =\hat{J}^T \in\hR^{3 \times 3}$ and $\hat{\btau}_d$ the estimates for the true spacecraft inertia $J$ and external disturbance torque $\btau_d$, respectively, and they can be obtained from system modeling. The corresponding estimation errors are $\tilde{J} = \hat{J} - J$ and $\tilde{\btau}_d= \hat{\btau}_d - \btau_d$. Similarly, denote by $\hat{E}(t)=\textup(\hat{e}_1(t), \cdots, \hat{e}_m(t)) \in\hR^{m\times m}$, $0\leq \hat{e}_i(t) \leq 1$, $i=1, \cdots, m$, an estimate of the actuator health indicator matrix which can be obtained from a real-time fault diagnosis algorithm. The estimation error is given by $\tilde{E}(t) = E(t) - \hat{E}(t)$.

For the purpose of controller development, $\hat{J}$, $\hat{\btau}_d$, $\hat{E}(t)$, $\bcq_d$, $\bomega_d$, and $\dot{\bomega}_d$ are assumed to be known. The following assumptions are required and reasonable for realistic spacecraft.

\begin{assumption} \label{assum2}
The spacecraft is fully actuated despite actuator faults or failures of some actuators, i.e., $\textup{rank}(E(t)) = \textup{rank}(\hat{E}(t)) = 3$. Furthermore, the estimated health indicator $\hat{E}(t)$ and its discrepancy $\tilde{E}(t)$ satisfy $  \| H \| \leq \rho_E$, where $H \triangleq  D \tilde{E}(t) \hat{E}^2(t) D^T (D \hat{E}^3(t) D^T)^{-1}$ and $ 0 \leq \rho_E <1$ is a constant.
\end{assumption}

\begin{assumption} \label{assum3}
$\tilde{J}$, $\tilde{\btau}_d$ and $\hat{\btau}_d$ are bounded by some constants $\rho_J, \rho_d, \hat{\rho}_d  \geq 0$ as $\|\tilde{J}\| \leq \rho_J$, $\|\tilde{\btau}_d\| \leq \rho_d$, and $\|\hat{\btau}_d\| \leq \hat{\rho}_d$ .
\end{assumption}

\begin{assumption} \label{assum4}
There exist known constants $\lambda_r \geq \lambda_l>0$ such that $\lambda_l \leq \lambda_{\min}(J) \leq \lambda_{\max}(J)\leq \lambda_r$.
\end{assumption}

\begin{assumption} \label{assum5}
$\bomega_d(t)$ and $\dot{\bomega}_d(t)$ are continuous functions of time and bounded by some constants $\rho_v, \rho_a\geq 0$ as $\|\bomega_d(t)\| \leq \rho_v$ and $\|\dot{\bomega}_d(t)\| \leq \rho_a$.
\end{assumption}

\begin{remark}
Assumption~\ref{assum2} is also utilized in \cite{Shen:15AUTO} and implies that the actuators can always provide three independent control torques irrespective of the possible actuator faults or failures, and the discrepancy between $E(t)$ and $\hat{E}(t)$ cannot be arbitrarily large.
\end{remark}

\begin{lemma}\cite{deRuiter:13} \label{lem1}
  Let $\bsr = \bomega_e + \Lambda\bsq_{e}$, $\Lambda = \Lambda^T>0$. If there exists an $\bar{r} \geq 0$ and a finite $T \geq 0$ such that $\|\bsr(t)\| \leq \bar{r}$ for all $t \geq T$, then $\limsup_{t\to\infty} \|\bsq_{e}(t)\| \leq \bar{r}/\lambda_{\min}(\Lambda)$ and $\limsup_{t\to\infty} \|\bomega_{e}(t)\| \leq (\lambda_{\max}(\Lambda) / \lambda_{\min}(\Lambda) + 1)\bar{r}$
\end{lemma}

\section{Main Results}

\subsection{Control Law Design with Estimated States}
By means of the estimated states $\hat{Q}$ and $\hat{\bomega}$, the estimated attitude and angular velocity tracking errors are computed as

\begin{equation*}
\begin{array}{c}
\hat{\bcq}_e = \bcq_d^{-1} \otimes \hat{\bcq}  \\
\hat{\bomega}_e = \hat{\bomega} - R(\hat{\bcq}_e)\bomega_d = \hat{\bomega} - \hat{\bar{\bomega}}_d, \quad \hat{\bar{\bomega}}_d \triangleq R(\hat{\bcq}_e)\bomega_d
\end{array}
\end{equation*}
Recalling $\bcq_e = \bcq_d^{-1} \otimes \bcq$ and $\hat{\bcq} = \bcq \otimes \tilde{\bcq}^{-1}$, one can obtain $\hat{\bcq}_e = \bcq_e \otimes \tilde{\bcq}^{-1}$ and thus

\begin{equation}\label{eq:hat:qe}
\hat{\bcq}_e = \bcq_e + M(\tilde{\bcq})\bcq_e, \quad \hat{\bsq}_e = \bsq_e + E(\tilde{\bcq})\bcq_e
\end{equation}
where $E(\tilde{\bcq}) = \left[-\tilde{\bsq} \,\,\,\, (\tilde{q}_0 - 1)I_3 + \tilde{\bsq}^{\times}\right]$ and

\begin{equation*}
\begin{array}{c}
M(\tilde{\bcq}) = \left[\begin{array}{cc}
 \tilde{q}_0 - 1 & \tilde{\bsq}^T \\
  -\tilde{\bsq} & (\tilde{q}_0 - 1)I_3 + \tilde{\bsq}^{\times}
\end{array}\right]
\end{array}
\end{equation*}
Note that $\hat{\bomega}_e$ can also be written as \cite{Gui:18JGCD2}

\begin{equation}\label{eq:hat:we}
\hat{\bomega}_e = \bomega_e + \bse_{\omega}, \quad \bse_{\omega} = (I_3 - R^T(\tilde{\bcq}))\bar{\bomega}_d + \tilde{\bomega}
\end{equation}
Additionally, the estimated sliding variable is computed by

\begin{equation}\label{eq:hat:s}
\begin{array}{c}
  \hat{\bss} = \hat{\bomega}_e + k\hat{\bsq}_{e} = \bss + \bse_s \\
  \bse_s = \tilde{\bomega} + (I_3 - R^T(\tilde{\bcq}))\bar{\bomega}_d + k E(\tilde{\bcq}) \bcq_e
\end{array}
\end{equation}

Substituting~(\ref{eq:tauc}) into (\ref{eq:dot:s1}) and simple algebraic manipulations lead to

\begin{equation}\label{eq:dot:s2}
\begin{array}{lll}
J\dot{\bss} & = & \Xi(J, \bomega_e, \bar{\bomega}_d)\bss + \frac{1}{2}k [\bsq_{e}^{\times} J + J \bsq_{e}^{\times}] \bss  \\
  & & + \hat{\bpsi} - \hat{\bpsi}_d + \hat{\btau}_d + \btau_r + D E(t) \btau_u
\end{array}
\end{equation}
where

\begin{equation}\label{eq:hat:psid}
\hat{\bpsi}_d = \hat{\bar{\bomega}}_d^{\times} \hat{J} \hat{\bar{\bomega}}_d + \hat{J} R(\hat{\bcq}_e)\dot{\bomega}_d
\end{equation}

\begin{equation}\label{eq:hat:psi}
\hat{\bpsi} = - \frac{1}{2}k^2 \hat{\bsq}_e^{\times} \hat{J} \hat{\bsq}_e + \frac{1}{2}k G(\hat{\bcq}_e) \hat{J} \hat{\bomega}_e - k \Xi(\hat{J}, 0, \hat{\bar{\bomega}}_d)\hat{\bsq}_{e}
\end{equation}

\begin{equation}\label{eq:taur}
\btau_r = \bpsi - \hat{\bpsi} + \hat{\bpsi}_d - \bpsi_d + \btau_d - \hat{\btau}_d
\end{equation}

Evidently, the term $\hat{\bpsi} - \hat{\bpsi}_d + \hat{\btau}_d$ can be fully canceled by feedback when actuators are healthy. The term $\btau_r$ represents the system uncertainties excluding actuator faults, and its upper bound is computed in the following.

First, it can be computed that

\begin{equation} \label{eq:id1}
\begin{array}{l}
\bpsi - \hat{\bpsi} = -\frac{1}{2}k^2 (E(\tilde{\bcq})\bcq_e)^{\times} \hat{J} \bsq_e - \frac{1}{2}k^2 \hat{\bsq}_{e}^{\times}  \hat{J} E(\tilde{\bcq}) \bcq_e   \\
\quad - \frac{1}{2} k^2  \bsq_e^{\times} \tilde{J}  \bsq_e - \frac{1}{2}k G(\hat{\bcq}_e) \hat{J} \bse_{\omega} - \frac{1}{2}k G(\bcq_e) \tilde{J} \bomega_e  \\
 \quad  - \frac{1}{2}k G(M(\tilde{\bcq})\bcq_e) \hat{J} \bomega_e + k \Xi(\tilde{J}, 0, \bar{\bomega}_d) \bsq_{e}  \\
 \quad  + k \Xi(\hat{J}, 0, \hat{\bar{\bomega}}_d) E(\tilde{\bcq})\bcq_e + k \Xi(\hat{J}, 0,  \bse_{\omega d}) \bsq_{e}
\end{array}
\end{equation}
where $\hat{\bar{\bomega}}_d = \bar{\bomega}_d + \bse_{\omega d}$ and $\bse_{\omega d} = (R^T(\tilde{\bcq}) - I_3) \bar{\bomega}_d$ are utilized. As shown in \cite{Gui:18JGCD2}, the inequality $\|I_3 - R^T(\tilde{\bcq})\| \leq 2\|\tilde{\bsq}\|$ holds. Note that $M^T (\tilde{\bcq}) M(\tilde{\bcq}) = 2(1 - \tilde{q}_0) I_4$ and $E^T(\tilde{\bcq}) E(\tilde{\bcq}) = 2(1 - \tilde{q}_0) I_3$, which implies $\|M(\tilde{\bcq})\| = \|E(\tilde{\bcq})\| =  \sqrt{2(1 - \tilde{q}_0)}$. Denoting $\rho_0 \triangleq (2(1 - \sqrt{1 - \rho_q^2}))^{\frac{1}{2}}$, Assumption~\ref{assum1} can then be utilized to verify that

\begin{equation}\label{eq:id2}
\limsup\limits_{t\to\infty} \|I_3 - R^T(\tilde{\bcq}(t))\| \leq 2\rho_q
\end{equation}

\begin{equation}\label{eq:id3}
\begin{array}{c}
\limsup\limits_{t\to\infty}\|M(\tilde{\bcq}(t))\| = \limsup\limits_{t\to\infty}\|E(\tilde{\bcq}(t))\|  \leq  \rho_0
\end{array}
\end{equation}

By means of Assumptions~\ref{assum2}-\ref{assum5} and (\ref{eq:id1})-(\ref{eq:id3}) and applying Cauchy-Schwartz inequality when appropriate, one can show that

\begin{equation} \label{eq:bd:es}
\limsup\limits_{t\to\infty} \| \bse_s(t) \| \leq \rho_s \triangleq \rho_{\omega} + 2 \rho_q\rho_v + k\rho_0
\end{equation}

\begin{equation*}
\begin{array}{c}
\limsup\limits_{t\to\infty} \| \hat{\psi}_d(t) - \psi_d(t) \| \leq  4\rho_q \rho_v^2 \|\hat{J}\| + 2 \rho_q\rho_a \|\hat{J}\| \\
  \quad \quad \quad \quad \quad \quad \quad \quad  + \rho_J \rho_v^2 + \rho_J \rho_a \\
\end{array}
\end{equation*}

\begin{equation*}
\limsup\limits_{t\to\infty} \| \btau_d(t) - \hat{\btau}_d(t) \| \leq  \rho_d
\end{equation*}

\begin{equation*}
\begin{array}{l}
\limsup\limits_{t\to\infty} \| \psi(t) - \hat{\psi}(t) \| \leq \frac{1}{2} k (\rho_0\|\hat{J}\| + \rho_J) \| \bomega_e \|  \\
  \quad   + \frac{1}{2} k^2 \rho_J \| \bsq_e \|^2 + (k^2\rho_0\|\hat{J}\|  + 3 k \rho_v(\rho_J + 2\rho_q\|\hat{J}\|)) \| \bsq_e\| \\
 \quad  \frac{1}{2}k^2 \rho_0^2 \|\hat{J}\|  +  \frac{1}{2}k (\rho_{\omega} + 2 \rho_q\rho_v) \|\hat{J}\| + 3k \rho_v \rho_0 \|\hat{J}\|
\end{array}
\end{equation*}
where $\bomega_e$ and $\bsq_e$ are viewed as fixed values when deriving the above equations. Noting $\bomega_e = \bss - k \bsq_e$, we can further derive

\begin{equation}\label{eq:bd:taur}
\limsup_{t\to\infty} \|\btau_r(t)\| \leq a_3 \|\bss\| + a_2 \|\bsq_{e}\|^2  + a_1 \|\bsq_{e}\| + a_0
\end{equation}
where $\bss$ and $\bsq_e$ are viewed as fixed values when deriving (\ref{eq:bd:taur}), and
\begin{equation}\label{eq:a2}
a_3 = \frac{1}{2} k (\rho_0\|\hat{J}\| + \rho_J), \quad  a_2 = \frac{1}{2} k^2 \rho_J
\end{equation}

\begin{equation}\label{eq:a1}
a_1 =  k^2\rho_0\|\hat{J}\| + k a_3 + 3 k \rho_v(\rho_J + 2\rho_q\|\hat{J}\|)
\end{equation}

\begin{equation}\label{eq:a0}
\begin{array}{l}
 a_0 = \frac{1}{2}k^2 \rho_0^2 \|\hat{J}\|  +  \frac{1}{2}k (\rho_{\omega} + 2 \rho_q\rho_v) \|\hat{J}\|    \\
  \quad \quad + 3k \rho_v \rho_0 \|\hat{J}\|  + 4\rho_q \rho_v^2 \|\hat{J}\| + 2 \rho_q\rho_a \|\hat{J}\|  \\
  \quad \quad   + \rho_J \rho_v^2 + \rho_J \rho_a + \rho_d
\end{array}
\end{equation}

An attitude tracking controller is designed as

\begin{equation} \label{eq:tauu}
\btau_u =\hat{E}^2(t) D^T (D \hat{E}^3(t) D^T)^{-1} \bsu
\end{equation}

\begin{equation}\label{eq:u}
\bsu = -K \hat{\bss} + \bsu_s + \hat{\bpsi}_d - \hat{\bpsi} - \hat{\btau}_d
\end{equation}

\begin{equation}\label{eq:us}
\bsu_s = \left\{\begin{array}{ll}
           -\frac{a_1 (\|\hat{\bsq}_{e}\| + \gamma) + a_0}{\| \hat{\bss} \|} \hat{\bss},  & \quad  \|\hat{\bss} \| \geq \varepsilon \\
            -\frac{a_1 (\|\hat{\bsq}_{e}\| + \gamma) + a_0}{\varepsilon} \hat{\bss}, & \quad \|\hat{\bss} \| < \varepsilon
          \end{array}\right.
\end{equation}
where $K=K^T \in\hR^{3\times3}$ is a positive definite matrix and $\gamma>0$ is a constant to be properly selected. The torque allocation algorithm given in~(\ref{eq:tauu}) was first proposed in \cite{Shen:15AUTO} and shown to minimize the cost $\btau_u^T \hat{E}^{-1}(t) \btau_u$ with the constraint $D \hat{E}(t) \btau_u = \bsu$. It delivers the virtual control torque $\bsu$ to each actuator in a manner that minimizes the use of faulty actuators. $\bsu_s$ aims to compensate for the influence of $\btau_r$, and, together with $-K \hat{\bss}$, stabilizes the sliding variable toward zero. Clearly, the control law is continuous at the switching surface $\hat{\bss} = \varepsilon$. Given the magnitude of estimation errors associated with $\hat{\bss}$, it is reasonable to select $\varepsilon > \rho_s$. The closed-loop stability under the above control law is analyzed in the next section.

\subsection{Stability Analysis}

\begin{theorem} \label{thm1}
Suppose that the spacecraft attitude and angular velocity estimates are generated by an observer satisfying Assumption~\ref{assum1}. Consider the attitude tracking system given by~(\ref{eq:dot:qe}), (\ref{eq:dot:s1}), and (\ref{eq:tauu})-(\ref{eq:us}) with $k >0$, $\lambda_{\min}(K) > a_3 + \rho_E(k \|\hat{J}\|/2 + \lambda_{\max}(K))$, $\gamma>0$, and $\varepsilon > \rho_s$. Then, the system tracking errors are ultimately bounded by

\begin{equation} \label{eq:si}
\limsup_{t\to\infty}\|\bss(t)\|
 \leq \bar{s}_{i} \triangleq \sqrt{\frac{\lambda_r}{\lambda_l}} \frac{\bar{\phi}(\bar{q}_{i-1}, 0)}{\kappa}
\end{equation}

\begin{equation} \label{eq:qi}
\limsup_{t\to\infty}\|\bsq_e(t)\|
 \leq \bar{q}_i, \quad \limsup_{t\to\infty}\|\bomega_e(t)\|
 \leq  2\bar{s}_i
\end{equation}
where $\bar{q}_i \triangleq k^{-1} \bar{s}_i$, $i\in\hZ_{+}$, $\bar{q}_0 = 1$, $\kappa$ is a positive constant given in (\ref{eq:kappa}), and $\bar{\phi}(\cdot, \cdot)$ is given in~(\ref{eq:bar:phi}). Additionally, if $0 \leq \bar{q}_1 <1$, both sequences $\{ \bar{s}_i\}_{i\in\hZ_+}$ and $\{ \bar{q}_i\}_{i\in\hZ_+}$ are strictly decreasing and convergent.
\end{theorem}

\begin{IEEEproof}
Substituting (\ref{eq:tauu}) and (\ref{eq:u}) into (\ref{eq:dot:s2}) and recalling $H =  D \tilde{E}(t) \hat{E}^2(t) D^T  (D \hat{E}^3(t) D^T)^{-1}$ leads to

\begin{equation}\label{eq:dot:s3}
\begin{array}{lll}
J\dot{\bss} & = & \Xi(J, \bomega_e, \bar{\bomega}_d)\bss + \frac{1}{2}k [\bsq_{e}^{\times} J + J \bsq_{e}^{\times}] \bss  \\
  & & - K \hat{\bss} + \bsu_s +  \btau_r + H \bsu
\end{array}
\end{equation}

First, we need to show that there is no finite escape time for the closed-loop system. Consider the following Lyapunov function candidate

\begin{equation}\label{eq:V}
V = \frac{1}{2} \bss^T J \bss
\end{equation}
Taking its time derivative along~(\ref{eq:dot:s3}) yields

\begin{equation} \label{eq:dot:V1}
  \dot{V} = \bss^T J \dot{\bss} = \bss^T( - K \hat{\bss} + \bsu_s +  \btau_r + H \bsu )
\end{equation}
where $\bss^T\Xi(J, \bomega_e, \bar{\bomega}_d)\bss = 0$ and $ \bss^T [\bsq_{e}^{\times} J + J \bsq_{e}^{\times}] \bss =0$ are utilized in the above derivations.

By analyzing the upper bounds of $\hat{\bss}$, $\bsu_s$, $\btau_r$, and $H \bsu$ and invoking Assumptions~\ref{assum1}-\ref{assum5}, it is readily to conclude that there exists finite constants $\eta_0, \eta_1>0$ such that $  \dot{V} \leq \eta_1 \|\bss\|^2/2 + \eta_0$. Noting

\begin{equation} \label{eq:bd:V}
\lambda_l \|\bss\|^2 \leq 2V \leq \lambda_r \|\bss\|^2
\end{equation}
it follows that $\dot{V} \leq \eta_1 V/\lambda_l + \eta_0 $. Applying the comparison principle \cite{Khalil:02} amounts to $ V(t) \leq e^{\eta_1 t/\lambda_l }(V(0) + \lambda_l \eta_0/\eta_1) - \lambda_l \eta_0/\eta_1$. Therefore,  $\bss(t)$ remains bounded in finite time. Note that $\bcq_e(t)$ is trivially bounded. Invoking Lemma~\ref{lem1}, it follows that $\bomega_e(t)$ remains bounded as well. Hence, the closed-loop system cannot escape in finite time. The remaining proof is divided into two steps.

\textit{Step 1) Preliminary Lyapunov Analysis:} Note that

\begin{equation} \label{eq:bd:u1}
\begin{array}{lll}
 - \bss^T K \hat{\bss} & = & -\bss^T K \bss - \bss^T K \bse_s \\
& \leq &  - \lambda_{\min}(K) \|\bss \|^2 + \lambda_{\max}(K) \|\bse_s\| \|\bss \|
\end{array}
\end{equation}
Additionally, if $\|\hat{\bss} \| \geq \varepsilon$, we have $- \bss^T(\bss + \bse_s ) = - \| \bss \|^2 - \bss^T \bse_s  \leq  - \| \bss \| (\| \hat{\bss} \| - \|\bse_s\|) +  \| \bss \|  \|\bse_s\| = -\| \bss \| \| \hat{\bss} \| + 2 \| \bss \|  \|\bse_s\| $. It then follows from~(\ref{eq:us}) that

\begin{equation}  \label{eq:bd:u2}
\begin{array}{lll}
\bss^T \bsu_s & = & -\frac{a_1 (\|\hat{\bsq}_{e}\| + \gamma) + a_0}{\| \hat{\bss} \|} \bss^T(\bss + \bse_s ) \\
   & \leq & - [a_1 (\|\hat{\bsq}_{e}\| + \gamma) + a_0]  \|\bss\| \\
   & & + 2[a_1 (\|\hat{\bsq}_{e}\| + \gamma) + a_0] \frac{\|\bse_s\|}{\hat{\|\bss\|}} \|\bss\|  \\
   & \leq & - [a_1 (\|\bsq_{e}\| + \gamma - \|E(\tilde{\bcq})\|) + a_0]  \|\bss\| \\
   & & + 2\varepsilon^{-1} [a_1 (\|\bsq_e\| + \gamma + \|E(\tilde{\bcq})\|) + a_0] \|\bse_s\|\|\bss\|  \\
\end{array}
\end{equation}
where $\|\hat{\bss} \| \geq \varepsilon$, $ \|\bsq_e\| - \|E(\tilde{\bcq})\| \| \bcq_e\| \leq  \|\hat{\bsq}_e\| \leq \|\bsq_e\| + \|E(\tilde{\bcq})\| \| \bcq_e\|$, and $\|\bcq_e\|=1$ are utilized in the above derivations. If $\|\hat{\bss}\| < \varepsilon$,  (\ref{eq:us}) can be used to verify

\begin{equation}  \label{eq:bd:u3}
\begin{array}{lll}
\bss^T \bsu_s &  = & -\frac{a_1 (\|\hat{\bsq}_{e}\| + \gamma) + a_0}{\varepsilon} \bss^T (\bss + \bse_s ) \\
   & \leq & -\frac{a_1 (\|\hat{\bsq}_{e}\| + \gamma) + a_0}{\varepsilon} \|\bss\|^2 + \frac{a_1 (\|\hat{\bsq}_{e}\| + \gamma) + a_0}{\varepsilon} \|\bse_s\| \|\bss\| \\
   & \leq &  \varepsilon^{-1} [a_1 (\|\bsq_{e}\| + \gamma + \|E(\tilde{\bcq})\|) + a_0] \|\bse_s\| \|\bss\|
\end{array}
\end{equation}
In addition, it can be shown that

\begin{equation}\label{eq:bd:u4}
\limsup_{t\to\infty} \|H (t) \bsu(t)\| \leq \rho_E ( b_3 \|\bss\| +  b_2 \|\bsq_{e}\|^2 +  b_1 \|\bsq_{e}\| +  b_0 )
\end{equation}
where $\bss$ and $\bsq_e$ are viewed as fixed values when deriving (\ref{eq:bd:u4}), and
\begin{equation*}\label{eq:b2}
b_3 = \frac{1}{2} k \|\hat{J}\| + \lambda_{\max}(K), \quad b_2 = \frac{1}{2} k^2  \|\hat{J}\|
\end{equation*}

\begin{equation*}\label{eq:b1}
b_1 = 2 b_2 \rho_0 + \frac{1}{2} k^2  \|\hat{J}\| + 3 k \rho_v \|\hat{J}\| + a_1
\end{equation*}

\begin{equation*}\label{eq:b0}
\begin{array}{l}
b_0 = b_2 \rho_0^2  + \frac{1}{2}k (\rho_w + 2\rho_q \rho_v) \|\hat{J}\| + 3 k \rho_v \rho_0 \|\hat{J}\|  \\
\quad \quad + \lambda_{\max}(K) \rho_s + a_1 (\rho_0 + \gamma) + a_0 + (\rho_v^2 + \rho_a )\|\hat{J}\| + \hat{\rho}_d
\end{array}
\end{equation*}

We then have the following lemma.

\begin{lemma} \label{lem2}
For arbitrarily small $\delta>0$, there exists $t_{\delta}>0$ such that the following inequalities

\begin{equation*}
\|E(\tilde{\bcq})\|  \leq  \rho_0 + \delta, \quad \| \bse_s \| \leq \rho_s + \delta
\end{equation*}

\begin{equation*}
 \|\btau_r\| \leq a_3 \|\bss\| + a_2 \|\bsq_{e}\|^2 + a_1 \|\bsq_{e}\| + a_0 + \delta
\end{equation*}

\begin{equation*}
\|H \bsu\| \leq \rho_E b_3 \|\bss\| + \rho_E b_2 \|\bsq_{e}\|^2 + \rho_E b_1 \|\bsq_{e}\| + \rho_E b_0 + \delta
\end{equation*}
hold for all $t\geq t_{\delta}$.
\end{lemma}

Employing these bounds given in Lemma~\ref{lem2} and combining~(\ref{eq:bd:u1})-(\ref{eq:bd:u4}), it can be obtained from~(\ref{eq:dot:V1}) that

\begin{equation} \label{eq:dot:V2}
\left\{\begin{array}{l}
  \dot{V} \leq -\kappa\|\bss\|^2 + \phi_1(\|\bsq_e\|, \delta) \|\bss\|, \quad \|\hat{\bss}\| \geq \varepsilon \\
  \dot{V} \leq -\kappa\|\bss\|^2 + \phi_2(\|\bsq_e\|, \delta) \|\bss\|, \quad \|\hat{\bss}\| < \varepsilon
\end{array}\right.
\end{equation}
holds for all $t\geq t_{\delta}$, where $\kappa$, $\phi_1(\cdot, \cdot)$ and $\phi_2(\cdot, \cdot)$ are defined as

\begin{equation} \label{eq:kappa}
\kappa = \lambda_{\min}(K) - a_3 - \rho_E b_3
\end{equation}

\begin{equation*} \label{eq:phi1}
\begin{array}{l}
 \phi_1(x, y) = (a_2 + \rho_E b_2)x^2 + [2\varepsilon^{-1} a_1 (\rho_s + y) + \rho_E b_1] x \\
\quad  + 2\varepsilon^{-1} (\rho_s + y) [a_1(\gamma + \rho_0 + y) + a_0] + \rho_E b_0 \\
\quad + (2 + \lambda_{\max}(K) + a_1) y - (a_1 \gamma - a_1 \rho_0 - \lambda_{\max}(K) \rho_s)
\end{array}
\end{equation*}

\begin{equation*} \label{eq:phi2}
\begin{array}{lll}
  \phi_2(x, y) = (a_2 + \rho_E b_2)x^2 + [\varepsilon^{-1} a_1 (\rho_s + y) + a_1 + \rho_E b_1] x \\
\quad + \varepsilon^{-1} (\rho_s + y) [a_1(\gamma + \rho_0 + y) + a_0] \\
 \quad + a_0 + \rho_E b_0 + \lambda_{\max}(K) (\rho_s + y) + 2y
\end{array}
\end{equation*}
for $x,y\in\hR$. Denote by
\begin{equation} \label{eq:bar:phi}
\bar{\phi}(\|\bsq_e\|, \delta) = \max\{\phi_1(\|\bsq_e\|, \delta), \phi_2(\|\bsq_e\|, \delta)\}
\end{equation}
It then follows from~(\ref{eq:bd:V}), (\ref{eq:dot:V2}) and (\ref{eq:bar:phi}) that

\begin{equation} \label{eq:dot:V3}
\begin{array}{l}
  \dot{V} \leq -\|\bss\| (\kappa\|\bss\| - \bar{\phi}(\|\bsq_e\|, \delta) ) \\
  \quad \leq -\|\bss\| (\kappa\sqrt{2V/\lambda_r} - \bar{\phi}(\|\bsq_e\|, \delta) )
\end{array} \quad\quad t\geq t_{\delta}
\end{equation}

\textit{Step 2) Sequential Lyapunov Analysis:} Next, an inductive argument is conducted by successively applying~(\ref{eq:dot:V3}). Note that $\|\bsq_e\| \leq \bar{q}_0 =1$ trivially holds. Since $\phi_i(\|\bsq_e\|, \delta)$, $i=1,2$, are both strictly increasing functions of $\|\bsq_e\|$, $\bar{\phi}(\|\bsq_e\|, \delta)$ is also a strictly increasing function of $\|\bsq_e\|$. Hence, $\bar{\phi}(\|\bsq_e\|, \delta) \leq \bar{\phi}(\bar{q}_0, \delta)$. Noting $\kappa>0$, (\ref{eq:dot:V3}) implies that $\dot{V} <0$ whenever $ \sqrt{2V} > \sqrt{\lambda_r} \bar{\phi}(\bar{q}_0, \delta) ) /\kappa$, and thus

\begin{equation*}
\limsup_{t\to\infty}\|\bss(t)\| \leq \limsup_{t\to\infty} \sqrt{\frac{2V(t)}{\lambda_l}}
 \leq \sqrt{\frac{\lambda_r}{\lambda_l}} \frac{\bar{\phi}(\bar{q}_0, \delta)}{\kappa}
\end{equation*}
Letting $\delta \to 0$ yields

\begin{equation*}
\limsup_{t\to\infty}\|\bss(t)\|
 \leq \bar{s}_1 \triangleq \sqrt{\frac{\lambda_r}{\lambda_l}} \frac{\bar{\phi}(\bar{q}_0, 0)}{\kappa}
\end{equation*}
Invoking Lemma~\ref{lem1} and the assumption that $0 \leq \bar{q}_1 <1$, it follows that $\limsup_{t\to\infty}\|\bsq_e(t)\| \leq  k^{-1} \bar{s}_1 =  \bar{q}_1 < \bar{q}_0 $ and $\limsup_{t\to\infty}\|\bomega_e(t)\| \leq 2\bar{s}_1$.

Next, assume that $0 \leq \bar{q}_i < \bar{q}_{i-1}$ and (\ref{eq:si}) and (\ref{eq:qi})  hold for $i\geq 1$. We intend to show that they also hold for $i+1$. From $\limsup_{t\to\infty} \|\bsq_e(t)\| \leq \bar{q}_i $, one can deduce that there exists $t_* \geq t_{\delta}$ such that $\|\bsq_e(t)\| \leq \bar{q}_{i} + \delta$ and $\bar{\phi}(\|\bsq_e(t)\|, \delta) \leq  \bar{\phi}(\bar{q}_{i} + \delta, \delta)$ hold for $t\geq t_*$. It then follows from (\ref{eq:dot:V3}) that

\begin{equation*}
\limsup_{t\to\infty}\|\bss(t)\| \leq \sqrt{\frac{\lambda_r}{\lambda_l}} \frac{\bar{\phi}(\bar{q}_i + \delta, \delta)}{\kappa}
\end{equation*}
which, by letting  $\delta \to 0$ , gives rise to

\begin{equation*}
\limsup_{t\to\infty}\|\bss(t)\|
 \leq \bar{s}_{i+1} \triangleq \sqrt{\frac{\lambda_r}{\lambda_l}} \frac{\bar{\phi}(\bar{q}_i, 0)}{\kappa}
\end{equation*}
By means of Lemma~\ref{lem1}, we have $\limsup_{t\to\infty}\|\bsq_e(t)\|
\leq k^{-1} \bar{s}_{i+1}=\bar{q}_{i+1}$ and $\limsup_{t\to\infty}\|\bomega_e(t)\| \leq 2\bar{s}_{i+1}$. Since $0\leq \bar{q}_i < \bar{q}_{i-1}$ and $\bar{\phi}(\cdot, 0)$ is a strictly increasing function, it follows that $0\leq \bar{s}_{i+1} < \bar{s}_i$ and $0\leq \bar{q}_{i+1} < \bar{q}_i$. This completes the inductive arguments and the results in Theorem~\ref{thm1} are now fully proven.
\end{IEEEproof}

As shown by Theorem~\ref{thm1}, the proposed controller can not only accommodate uncertain inertias, disturbance torques, state estimation errors, and actuator faults all together but hold a separation principle with a series of stable stand-alone observers. It admits computable, successively tighter predicted bounds on the ultimate tracking errors. These attractive properties together, according to the best of the author's knowledge, have never been achieved by existing attitude control laws. In contrast, the fault-tolerant attitude controllers derived in \cite{Shen:15AUTO, Gui:17RNC2, HuQL:19, Liu:18} and references therein assumed that the system states for control were precise, and there lacks a rigorous theory to address their stability when combined with observers/filters and their robustness against state estimation errors. In addition, the estimated bounds in \cite{Shen:15AUTO, Gui:17RNC2, HuQL:19} are not practically computable or give too conservative predictions because they rely on precise system parameters, state-dependent variables, or even undetermined parameters.

\subsection{Algorithms for Performance Bound Prediction}

Evidently,  $\{ \bar{q}_i\}_{i\in\hZ_+}$ and $\{ 2\bar{s}_i\}_{i\in\hZ_+}$ provide successively tighter upper bounds on the ultimate attitude and angular velocity tracking errors. Since both sequences are convergent, we can denote $\bar{q}_{\infty} = \lim_{i\to\infty} \bar{q}_{i}$ and $\bar{s}_{\infty} = \lim_{i\to\infty} \bar{s}_{i}$, and then have $\limsup_{t\to\infty}\|\bsq_e(t)\| \leq   \bar{q}_{\infty} $ and $\limsup_{t\to\infty}\|\bomega_e(t)\| \leq 2\bar{s}_{\infty}$. Additionally, $\|\hat{\bss}\| \leq \|\bss\| + \|\bse_s\|$ implies $\limsup_{t\to\infty} \|\hat{\bss}(t)\| \leq \bar{s}_{\infty} + \rho_s$.

Since we have chosen $\varepsilon > \rho_s$, it is possible that $\bar{s}_{\infty} + \rho_s < \varepsilon$. If $\bar{s}_{\infty} + \rho_s < \varepsilon$, there exists $t_{\varepsilon} \geq t_{\delta}$ such that $\|\hat{\bss}(t)\| < \varepsilon$ for all $t\geq t_{\varepsilon}$ and the closed-loop trajectory will enter the prespecified boundary layer. Consequently, an additional Lyapunov analysis can be conducted by taking the negative term $-\varepsilon^{-1}(a_1 (\|\hat{\bsq}_{e}\| + \gamma) + a_0)\|\bss\|^2$, as appeared in~(\ref{eq:bd:u3}) but omitted in the preceding analysis, into account so as to obtain much tighter performance bound prediction. One can derive from~(\ref{eq:bd:u3}) and (\ref{eq:dot:V2}) that

\begin{equation} \label{eq:dot:V4}
\begin{array}{l}
  \dot{V} \leq -(\kappa + \varepsilon^{-1}(a_1 (\|\hat{\bsq}_{e}\| + \gamma) + a_0))\|\bss\|^2 \\
  \quad \quad +  \phi_2(\|\bsq_e\|, \delta) \|\bss\| \\
\quad \leq -\|\bss\| (\kappa^{\prime} \|\bss\| - \phi_2(\|\bsq_e\|, \delta)) \\
\quad \leq -\|\bss\| (\kappa^{\prime} \sqrt{2V/\lambda_r} -  \phi_2(\|\bsq_e\|, \delta) )
\end{array}
\end{equation}
holds for $t\geq t_{\varepsilon}$, where $\kappa^{\prime} = \kappa + \varepsilon^{-1}(a_1 \gamma + a_0)$. Equation~(\ref{eq:dot:V4}) can be utilized to perform sequential Lyapunov analysis similarly to Step 2) in the proof of Theorem~\ref{thm1}. This yields another two nonnegative sequences $\{ \bar{q}_i ^{\prime} \}_{i\in\hZ_+}$ and $\{ \bar{s}_i^{\prime}\}_{i\in\hZ_+}$:

\begin{equation*}
\bar{q}_i^{\prime} = k^{-1}  \bar{s}_{i}^{\prime}, \quad \bar{s}_{i}^{\prime} = \sqrt{\frac{\lambda_r}{\lambda_l}} \frac{\phi_2(\bar{q}_{i-1}^{\prime}, 0)}{\kappa^{\prime}}, \quad \bar{q}_0^{\prime} = \bar{q}_{\infty}
\end{equation*}
which enables to obtain
\begin{equation*}
\limsup_{t\to\infty}\|\bss(t)\|
 \leq \bar{s}_{i}^{\prime}
\end{equation*}

\begin{equation*}
\limsup_{t\to\infty}\|\bsq_e(t)\|
 \leq \bar{q}_i^{\prime}, \quad \limsup_{t\to\infty}\|\bomega_e(t)\|
 \leq  2\bar{s}_i^{\prime}
\end{equation*}

Generally, $\bar{q}_{\infty} >0 $ when system uncertainties are nonzero. One can then derive $\bar{q}_1^{\prime} < \bar{q}_0^{\prime} = \bar{q}_{\infty}$ because $\bar{q}_1^{\prime} = k^{-1}  \bar{s}_{1}^{\prime}$, $\bar{q}_{\infty} = k^{-1}  \bar{s}_{\infty}$, $\kappa^{\prime} > \kappa$ and

\begin{equation*}
\bar{s}_{1}^{\prime}= \sqrt{\frac{\lambda_r}{\lambda_l}} \frac{\phi_2(\bar{q}_{0}^{\prime}, 0)}{\kappa^{\prime}}< \sqrt{\frac{\lambda_r}{\lambda_l}} \frac{\bar{\phi}(\bar{q}_{\infty}, 0)}{\kappa} = \bar{s}_{\infty}
\end{equation*}
Hence, the sequences  $\{ \bar{q}_i ^{\prime} \}_{i\in\hZ_+}$ and $\{ \bar{s}_i^{\prime}\}_{i\in\hZ_+}$ are both strictly decreasing and convergent, and provide tighter prediction on the ultimate performance bounds than $\{ \bar{q}_i\}_{i\in\hZ_+}$ and $\{ 2\bar{s}_i\}_{i\in\hZ_+}$. Letting $\bar{q}_{\infty}^{\prime} = \lim_{i\to\infty} \bar{q}_{i}^{\prime}$ and $\bar{s}_{\infty}^{\prime} = \lim_{i\to\infty} \bar{s}_{i}^{\prime}$, we have $\limsup_{t\to\infty}\|\bsq_e(t)\| \leq   \bar{q}_{\infty}^{\prime} $ and $\limsup_{t\to\infty}\|\bomega_e(t)\| \leq 2\bar{s}_{\infty}^{\prime}$.

Summarizing the above analysis, we can derive an algorithm for predicting ultimate bounds on the attitude and angular velocity tracking errors.

\begin{algorithm}
Set $\bar{q}_0 = 1$ and $i=1$, and select a small $\eta>0$.

\textbf{loop 1}

Compute $\bar{s}_i$ and $\bar{q}_i$.

If $|\bar{q}_i - \bar{q}_{i-1}| > \eta$, let $i = i+1$ and continue loop 1; otherwise, output $\bar{q}_{\infty}= \bar{q}_i$ and $\bar{s}_{\infty} = \bar{s}_i$, and stop loop 1.

\textbf{end loop 1}

If $\bar{s}_{\infty} + \rho_s < \varepsilon$, reset $i=1$ and $\bar{q}_0^{\prime} = \bar{q}_{\infty}$, and do loop 2.

\textbf{loop 2}

Compute $\bar{s}_i^{\prime}$ and $\bar{q}_i^{\prime}$.

If $| \bar{q}_i^{\prime} - \bar{q}_{i-1}^{\prime} | > \eta$, let $i = i+1$ and continue loop 2; otherwise, output $\bar{q}_{\infty}^{\prime} = \bar{q}_i^{\prime} $ and $\bar{s}_{\infty}^{\prime} = \bar{s}_i^{\prime}$, and stop loop 2.

\textbf{end loop 2}
\end{algorithm}

\begin{remark}
Note that observer-based adaptive controllers and PD controllers for robust attitude tracking were derived in \cite{deRuiter:16TAC} and \cite{Gui:18JGCD2} but both of them did not consider the influence of actuator faults. Additionally, the performance prediction algorithm derived in \cite{deRuiter:16TAC} needs not only the bounds on the attitude and angular velocity estimation errors but also the bounds on their rates, which are very difficult to obtain in practice. In contrast, the proposed performance prediction algorithm only requires the former and, in this sense, is much simpler to use.
\end{remark}

\section{Numerical Examples}

\begin{figure*}[!tbp]
\begin{center}
\includegraphics[height=6cm]{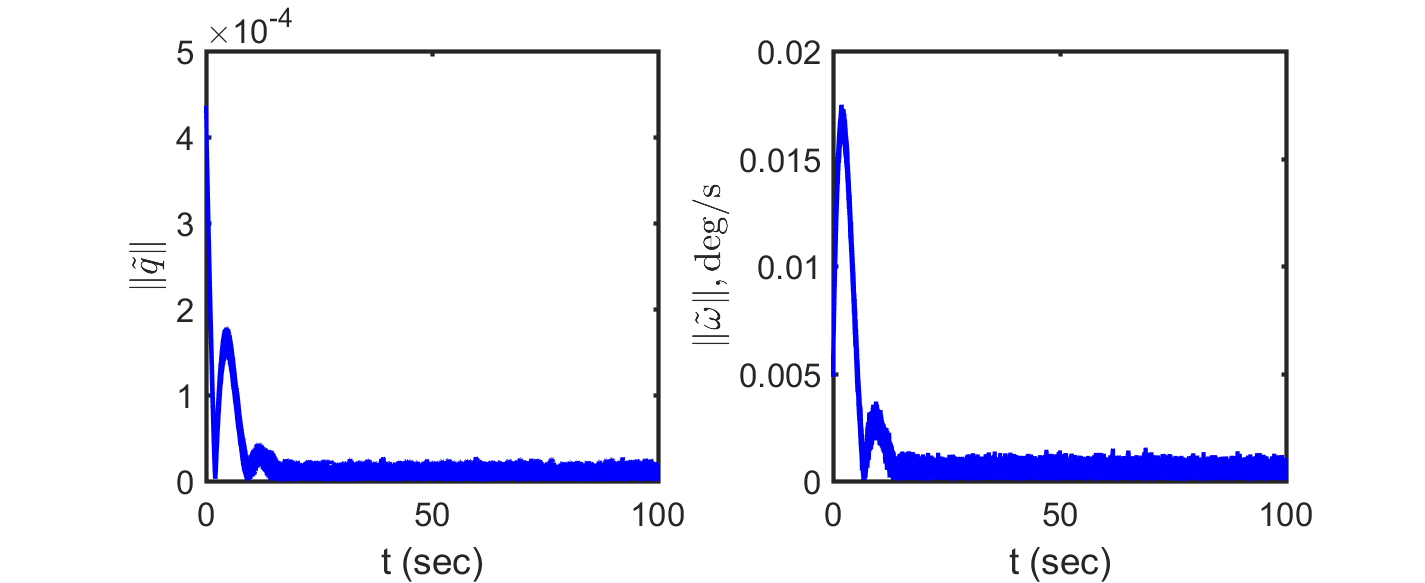}
\vspace{0 mm}
\caption{Attitude and angular velocity estimation errors by the observer in~\cite{Thienel:03}}  \label{fig:est_error}
\end{center}
\end{figure*}

\begin{figure*}[!tbp]
\begin{center}
{\includegraphics[height=6cm]{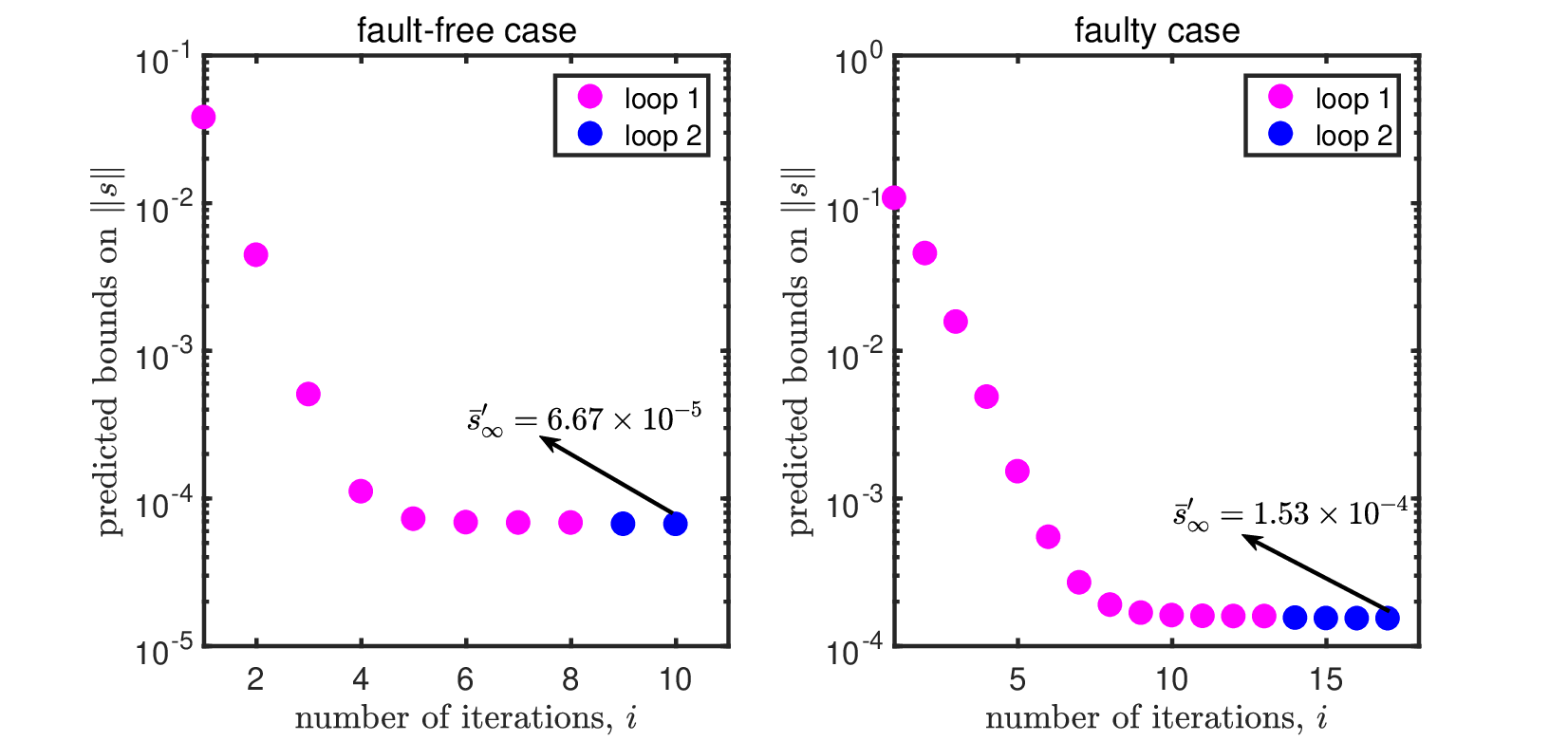}}
\vspace{0 mm}
\caption{Predicted ultimate bounds on $\|\bss\|$}  \label{fig:si}
\end{center}
\end{figure*}


\begin{figure*}[!tbp]
\begin{center}
{\includegraphics[height=6cm]{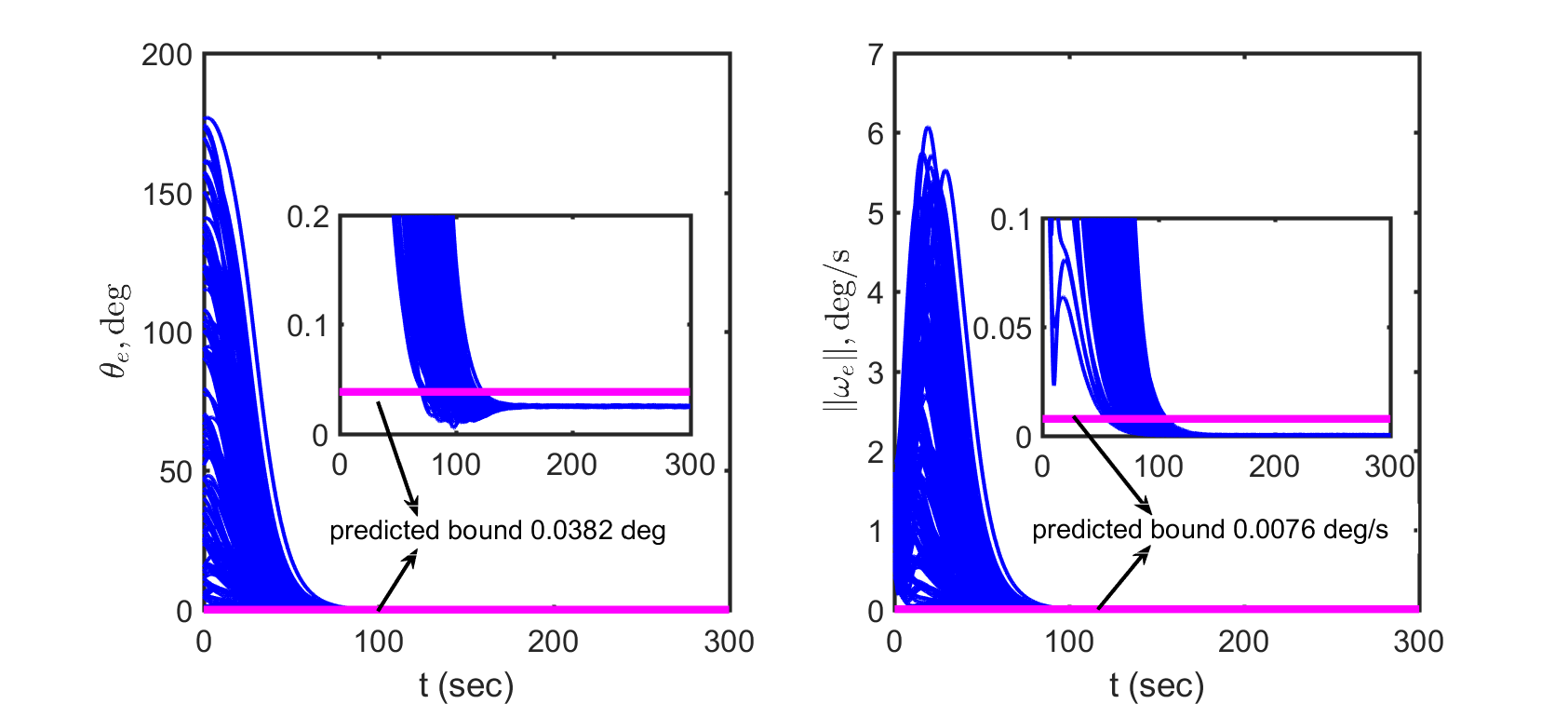}}
\vspace{0 mm}
\caption{Attitude and angular velocity tracking errors: fault-free case }  \label{fig:free}
\end{center}
\end{figure*}

\begin{figure*}[!tbp]
\begin{center}
{\includegraphics[height=8cm]{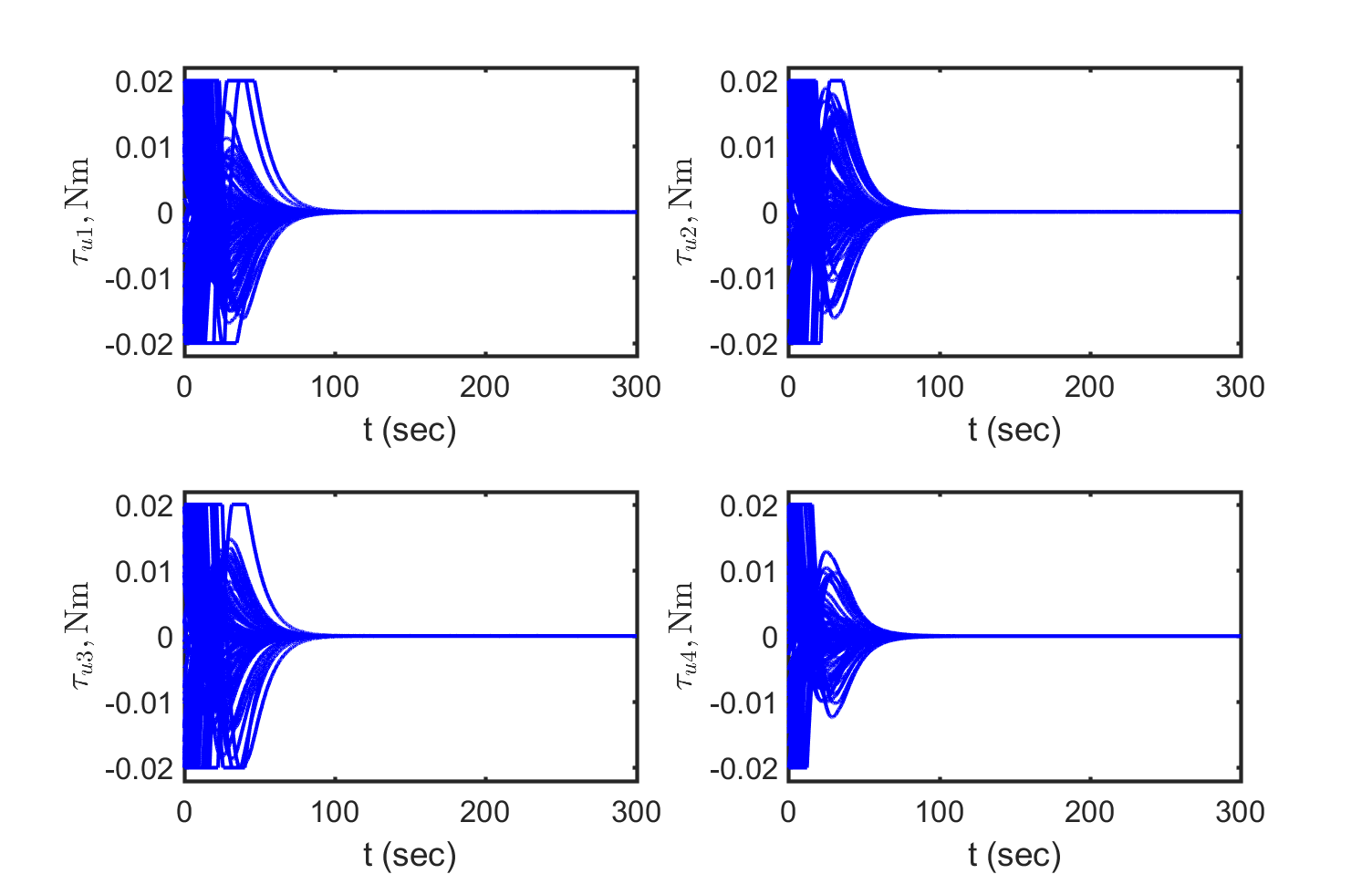}}
\vspace{0 mm}
\caption{Command torque fed to actuators: fault-free case }  \label{fig:tauu:free}
\end{center}
\end{figure*}

\begin{figure*}[!tbp]
\begin{center}
{\includegraphics[height=6cm]{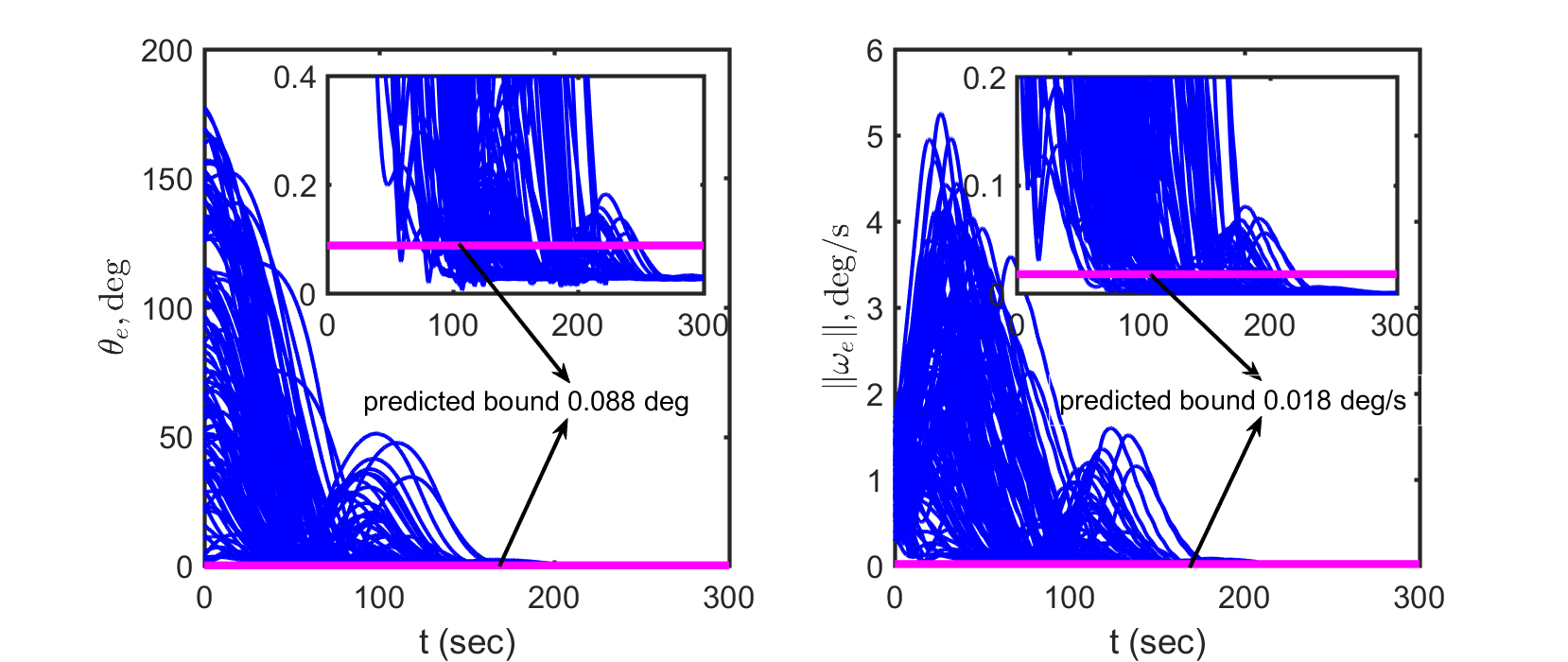}}
\vspace{0 mm}
\caption{Attitude and angular velocity tracking errors: faulty case }  \label{fig:fault}
\end{center}
\end{figure*}

\begin{figure*}[!tbp]
\begin{center}
{\includegraphics[height=8cm]{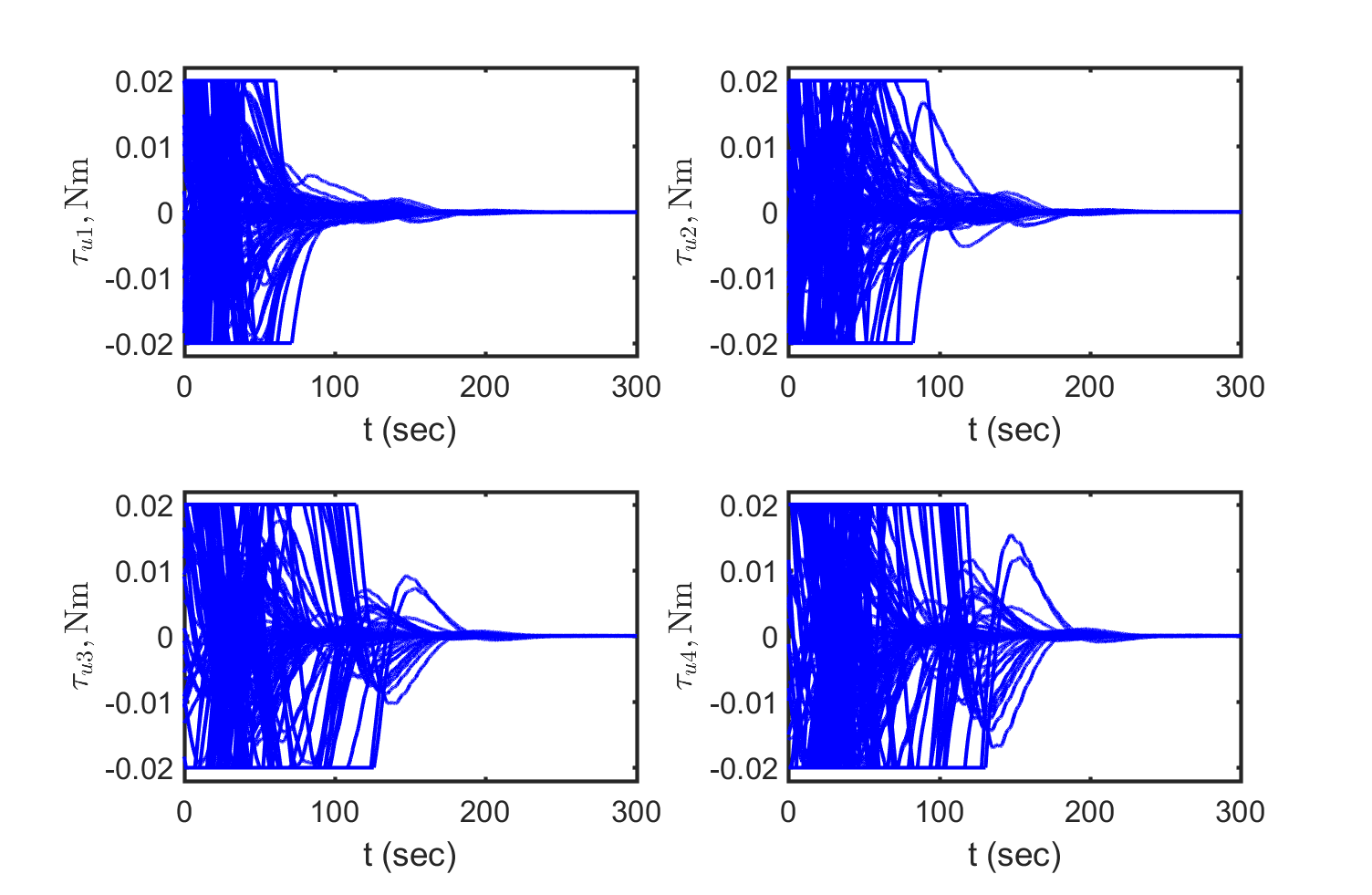}}
\vspace{0 mm}
\caption{Command torque fed to actuators: faulty case }  \label{fig:tauu:fault}
\end{center}
\end{figure*}

Next, the proposed method is demonstrated via numerical simulations. Consider a microsatellite with an inertia matrix of

\begin{equation*}
J = \left[\begin{array}{ccc}
         8 & 0.15 & -0.27 \\
         0.15 & 6.75 & -0.1 \\
         -0.27 & -0.1 & 6.25
       \end{array}\right] \, \mathrm{kg\cdot m^2}
\end{equation*}
The desired attitude trajectory is driven by $\bomega_d(t) = [2\cos(\omega_0 t), 2\sin(\omega_0 t), \sin(\omega_0 t)]^T \times 10^{-3} \,\mathrm{rad/s} $ with $\omega_0 = 1\times 10^{-3}$ \mbox{rad/s} and  $\bcq_d(0)=\vone $. The disturbance torque acting on the satellite is given by $ \btau_d = 2.5[\sin(\omega_0 t), -\cos(\omega_0 t), \cos(\omega_0 t)]^T \times 10^{-6}$ \mbox{Nm}. The estimated inertia and disturbance torque used for control are $\hat{J} = \mathrm{diag}\{8, 7, 6\} \,\mathrm{kg\cdot m^2}$ and $\hat{\btau}_d = 0$ \mbox{Nm}, respectively. Assume that the spacecraft carries four pairs of thrusters, whose torque distribution matrix is $\bcd = [1, 0, 0, 1/\sqrt{3}; 0, 1, 0, 1/\sqrt{3}; 0, 0, 1, 1/\sqrt{3}]$. The maximum output torque for each pair of thrusters is 0.02 \mbox{Nm}.


Note that the attitude quaternion $\bcq$ can also be written by means of the eigenaxis $\bsn$ ($\|\bsn\|=1$) and principal rotation angle $\theta$ as $\bcq= [\cos(\theta/2), \bsn^T \sin(\theta/2)]^T$. According to the Euler's rotation theorem,  the principal rotation angle $\theta$ is no more than 180 \mbox{deg}. The attitude and angular velocity measurements are generated by $\bcq_m = \bcq \otimes \tilde{\bcq}_m^{-1}$ and $\bomega_m = \bomega + \bsb + \betaa_u$, where $\tilde{\bcq}_m = [\cos(\tilde{\theta}_m), \tilde{\bse}_m^T \sin(\tilde{\theta}_m/2)]^T$ denotes the attitude measurement error, and $\tilde{\theta}_m$ follows the normal distribution $\mathcal{N}(0, (0.01 \,\mathrm{deg})^2 I_3)$ and $\tilde{\bse}_m$ is uniformly distributed on the two-dimensional unit sphere $\hS^2$; $\bsb$ is the gyro bias that satisfies $\dot{\bsb} = \betaa_v$ and $\bsb(0) = [-5, 15, -10]^T$ \mbox{deg/h}; and $\betaa_u$ and $\betaa_v$ are white Gaussian noise processes subject to the distributions $\mathcal{N}(0, (3\times 10^{-6} \,\mathrm{rad/s})^2 I_3)$ and $\mathcal{N}(0, (10^{-7} \mathrm{rad/s^{3/2}})^2 I_3)$, respectively.  $\bcq_m$ and $\bomega_m$ are fed to the global exponential observer derived in \cite{Thienel:03} (with $\hat{\bcq}(0)= \bcq_m(0)$, $\hat{\bsb}(0) = 0 $ \mbox{rad/s} and the observer gain $k_o = 1$) to provide $\hat{\bcq}$ and $\hat{\bomega} = \bomega_m - \hat{\bsb}$. The control parameters are selected as $k = 0.2$, $K = 0.7I_3$, $\varepsilon = 0.01$, and $\gamma = 0.01$. Monte Carlo simulations are performed first for the fault-free case, i.e., $\hat{E}=E=I_4$, and then for the case with  actuator faults satisfying $e_1(t) = 1 - 0.1|\sin t|$, $e_2(t) = 0.7 - 0.1\cos t$, $e_3(t) = 0$, $e_4(t) = 0.5 - 0.1 \sin t$, $\hat{e}_1(t) = \hat{e}_2(t) = 1 $, $\hat{e}_3 = 0$ and $\hat{e}_4(t) = 0.7$. Both cases are simulated for 100 instances with $\bomega(0)$ generated from a uniform distribution over $[-0.02, 0.02]$ \mbox{rad/s} and $\bcq(0)= [\cos(\theta(0)/2), \bsn^T(0) \sin(\theta(0)/2)]^T$, where  $\theta(0)$ is subject to a uniform distribution over $[0, \pi]$ \mbox{rad} and $\bsn(0)$ is uniformly distributed on $\hS^2$.

Figure~\ref{fig:est_error} plots the quaternion and angular velocity estimation errors during steady state for 100 simulation instances. Simple statistical analysis shows that $\limsup_{t\to\infty} \|\tilde{\bsq}\| \leq 2.15\times 10^{-5}$ and $\limsup_{t\to\infty} \|\tilde{\bomega}\| \leq 1.56\times 10^{-5}$ \mbox{rad/s}. Hence, we choose $\rho_q = 2.15\times 10^{-5}$ and $\rho_w = 1.56\times 10^{-5}$. Other parameters for performance bound prediction are set to $\rho_J = 0.5$, $\lambda_r = 8.5$, $\lambda_l = 6$, $\rho_v = 0.0022$, $\rho_a = 2.2\times 10^{-6}$, $\rho_d =\hat{\rho}_d = 3\times 10^{-6}$, $a_1 = 0.011$, $a_0 = 1.93\times 10^{-5}$, and $\rho_E = 0.08$. Algorithm 3.1 can then be utilized to compute the ultimate bounds on tracking errors recursively. Figure~\ref{fig:si} shows the predicted bounds on $\|\bss\|$ at each iteration for the fault-free case and faulty case, which converge to the limits of $\bar{s}_{\infty}^{\prime} = 6.67\times 10^{-5}$ and $\bar{s}_{\infty}^{\prime} = 1.53\times 10^{-4}$, respectively. The pink and blue dots represent the computed results by loops 1 and 2, respectively. For the fault-free case, Algorithm 3.1 keeps at loop 1 from the 1st iteration to the 8th iteration, and switches to the computation of loop 2 since the 9th iteration. The prediction converges to the specified threshold in 10 iterative computations. For the faulty case, the algorithm keeps at loop 1 from the 1st iteration to the 13th iteration, and switches to the computation of loop 2 since the 14th iteration. The prediction converges to the specified threshold in 17 iterative computations. For both cases, the predicted bounds become successively tighter and the prediction accuracy is improved dramatically (about three orders) compared to the first iteration, which verifies the effectiveness of Algorithm 3.1. In addition, Algorithm 3.1 significantly improves the prediction accuracy for the first few iterations. The improvements become minor for the last few iterations.

The blue curves in Fig.~\ref{fig:free} are the principal rotation angle of the attitude tracking error, $\theta_e$ ($\leq 180$ \mbox{deg} \cite{Sidi:00}), and the norm of the angular velocity tracking error, $\|\bomega_e\|$, for the 100 fault-free simulation instances. The predicted bounds are 0.0382 \mbox{deg} for $\theta_e$ and 0.0076 \mbox{deg/s} for $\|\bomega_e\|$ while the true steady-state errors are 0.027 \mbox{deg} for $\theta_e$ and $4.2\times10^{-4}$ \mbox{deg/s} for $\|\bomega_e\|$. The predicted bounds well envelope the true steady-state track errors but still have some conservativeness. This is due to the fact that we utilized the worst of all uncertainties for performance prediction. Figure~\ref{fig:tauu:free} plots the corresponding command torques computed by~(\ref{eq:tauu})-(\ref{eq:us}) and fed to actuators for the 100 simulated instances. It can be seen that the saturation limit is reached by many simulated instances at the beginning phase and left as tracking errors converge to the ultimate bounds.

Figure~\ref{fig:fault} presents the simulation results for the faulty case. The given fault profile implies that the third pair of thruster completely fails while the remaining three pairs suffer from fading output torques. In spite of this, the proposed controller exhibits significant fault-tolerance and stabilizes the tracking errors to $ |\theta_e| \leq 0.032 $ \mbox{deg} and $\|\bomega_e\| \leq 1.8 \times 10^{-3}$ \mbox{deg/s}. The actuator faults, however, induce more aggressive transient oscillations and longer convergence time compared to the fault-free case. In addition, the predicted ultimate bounds for $\theta_e$ and $\|\bomega_e\|$ are 0.0088 \mbox{deg} and 0.0227 \mbox{deg/s}, respectively, which become a bit more conservative than the fault-free case because the faults increase the system uncertainties. Figure~\ref{fig:tauu:fault} plots the corresponding command torques in the faulty case, which show more aggressive responses compared to the fault-free case due to the influence of actuator faults.

The performance of the proposed method is further compared with the adaptive fault-tolerant controller derived in~\cite{Shen:15AUTO} for initial conditions $\bcq_e = [ 0.7874, 0.2, -0.5, -0.3]^T$ and $\bomega_e = [0.02, 0.01, -0.025]^T$ \mbox{rad/s} and considering the above actuator faults profile. In order for a fair comparison, the control parameters of the two methods are tuned such that the convergence time for the attitude tracking error is about 150 seconds. The resultant control gains for the proposed method remain the same as given before.
The parameters used by~\cite{Shen:15AUTO} are $k=0.09$, $k_1 = k_2 = k_v = 0.1$, $\sigma = \alpha_1 = \alpha_2 = 0.01$, $\Delta_{\max} = 0.08$, $\hat{c}=0.01$, and $c_w = 0.0022$. The results are compared in Table~\ref{tab1} and both methods show significant robustness. The method in~\cite{Shen:15AUTO} attains slightly higher steady-state accuracy because its adaptive mechanism for uncertainty compensation. Note that the estimated performance bounds given in Theorem 1 of~\cite{Shen:15AUTO} require the true values of the inertia matrix and initial attitude and angular velocity tracking errors. Although these quantities are unknown for realistic applications, they are still assumed to be available so as to compute the bounds given in~\cite{Shen:15AUTO}. Despite using the aforementioned true quantities, the obtained performance prediction according to Theorem 1 of~\cite{Shen:15AUTO} is still quite poor. The predicted ultimate bounds are $1.52$ for $ \| \bsq_e \| $, which is actually unreasonable because the unit norm constraint of attitude quaternion already means $ \| \bsq_e \| \leq 1$, and $27.24$ \mbox{deg/s} for $ \|\bomega_e\| $, which deviates substantially from the true residual error $\|\bomega_e\| \leq 0.0014$  \mbox{deg/s}. In contrast, the proposed method achieves both satisfactory tracking control and far better performance prediction. In particular, the predicted bounds by the proposed algorithm are over $1\times 10^{4}$ times better than those obtained by~\cite{Shen:15AUTO}.


\begin{table}[!tbp]
\renewcommand{\arraystretch}{1.3}
\caption{Performance comparison for different methods}
\label{tab1}
\centering
\begin{tabular}{c|c|c|c|c}
\hline
controller & \multicolumn{2}{c|}{true residual error} &  \multicolumn{2}{c}{predicted bounds}  \\
\cline{2-5}
{} & $\|\bsq_e\|$ & $\|\bomega_e\|$,  \mbox{deg/s} & $\|\bsq_e\|$ & $\|\bomega_e\|$,  \mbox{deg/s} \\
\hline
proposed & $2.47\times 10^{-4}$  &  0.0018 &  $7.67\times 10^{-4}$  & 0.018 \\
\hline
Ref.~\cite{Shen:15AUTO} & $1.92\times 10^{-4}$ & 0.0014 & 1.52  &  27.24   \\
\hline
\end{tabular}
\end{table}


\section{Conclusion}

This paper presented a continuous sliding mode attitude controller that enjoys a separation principle with any stand-alone observer ensuring uniformly ultimately bounded estimation errors. Moreover, the controller can reject perturbations due to uncertain inertias, disturbance torques, state estimation errors, and actuator faults, and  stabilize the tracking error into a small neighborhood of zero. Sequential Lyapunov analysis techniques enables to derive an algorithm for computing successively tighter upper bounds on the ultimate state tracking errors. Our results not only provide a robust observer-based attitude tracking law but can also be utilized to assist in gain selection with guaranteed steady-state performance bounds without extensive Monte Carlo simulations.

\section*{Appendix A: Proof of~(\ref{eq:dot:s1})-(\ref{eq:psi})}

The derivations for~(\ref{eq:dot:s1})-(\ref{eq:psi}) are given in the following. First, note that $J\dot{\bomega}_e = [(J(\bomega_e + \bar{\bomega}_d))^{\times} - \bar{\bomega}_d^{\times}J - J\bar{\bomega}_d^{\times}] \bomega_e -  \bar{\bomega}_d^{\times}J \bar{\bomega}_d - J R(\bcq_e)\dot{\bomega}_d + \btau_c + \btau_d $ according to~\cite{Mayhew:11,Gui:17RNC1}. Applying the notations defined in~(\ref{eq:Xi}) and (\ref{eq:psi}), we have

\begin{equation*}
\begin{array}{lll}
J\dot{\bss} & = & J\dot{\bomega}_e +  k J \dot{\bsq}_{e}   \\
  & = & \Xi(J, \bomega_e, \bar{\bomega}_d)\bomega_e + \frac{1}{2}k J G(\bcq_e)  \bomega_e - \bpsi_d + \btau_c + \btau_d
\end{array}
\end{equation*}
Noting $\bomega_e = \bss - k \bsq_e$, we can derive

\begin{equation*}
\begin{array}{l}
\Xi(J, \bomega_e, \bar{\bomega}_d)\bomega_e + \frac{1}{2}k J G(\bcq_e)  \bomega_e \\
\quad = \Xi(J, \bomega_e, \bar{\bomega}_d)\bss - k \Xi(J, \bomega_e, \bar{\bomega}_d)\bsq_e + \frac{1}{2}k J G(\bcq_e)  \bomega_e
\end{array}
\end{equation*}
In addition, it follows that

\begin{equation*}
\begin{array}{l}
- k \Xi(J, \bomega_e, \bar{\bomega}_d)\bsq_e + \frac{1}{2}k J G(\bcq_e)  \bomega_e \\
\quad =  - k \Xi(J, 0, \bar{\bomega}_d)\bsq_e - k(J \bomega_e)^{\times} \bsq_e + \frac{1}{2}k J G(\bcq_e)  \bomega_e
\end{array}
\end{equation*}
and

\begin{equation*}
\begin{array}{l}
- k(J \bomega_e)^{\times} \bsq_e + \frac{1}{2}k J G(\bcq_e) \bomega_e \\
\quad =   k \bsq_e^{\times} J \bomega_e + \frac{1}{2}k J G(\bcq_e)  \bomega_e \\
\quad =   \frac{1}{2} k \bsq_e^{\times} J \bomega_e + \frac{1}{2} k \bsq_e^{\times} J \bomega_e + \frac{1}{2}k J \bsq_e^{\times}  \bomega_e + \frac{1}{2}k q_{e0} J  \bomega_e \\
\quad =   \frac{1}{2} k \bsq_e^{\times} J ( \bss - k \bsq_e ) + \frac{1}{2}k J \bsq_e^{\times} ( \bss - k \bsq_e ) + \frac{1}{2} k G(\bcq_e) J \bomega_e  \\
\quad =  \frac{1}{2} k [\bsq_e^{\times} J + J \bsq_e^{\times}] \bss - \frac{1}{2}k^2 \bsq_e^{\times} J \bsq_e + \frac{1}{2}k G(\bcq_e) J \bomega_e \\
\end{array}
\end{equation*}
Summarizing the above derivations, the expression of $J \dot{\bss}$ is given by

\begin{equation*}
\begin{array}{lll}
J\dot{\bss} & = & \Xi(J, \bomega_e, \bar{\bomega}_d)\bss + \frac{1}{2} k [\bsq_e^{\times} J + J \bsq_e^{\times}] \bss  \\
& & - \frac{1}{2}k^2 \bsq_e^{\times} J \bsq_e + \frac{1}{2}k G(\bcq_e) J \bomega_e - k \Xi(J, 0, \bar{\bomega}_d)\bsq_e \\
& & - \bpsi_d + \btau_c + \btau_d
\end{array}
\end{equation*}
Recalling $\bpsi$ defined in~(\ref{eq:psi}), one can then obtain~(\ref{eq:dot:s1}).

\section*{Appendix B: Proof of Lemma~\ref{lem2}}

Given arbitrarily small $\delta_i >0$, $i=1,2,3,4$, (\ref{eq:id3})-(\ref{eq:bd:taur}) and (\ref{eq:bd:u4}) imply that there exist $t_i \geq 0 $, $i=1,2,3,4$, such that the following inequalities hold

\begin{equation*}
\|E(\tilde{\bcq})\|  \leq  \rho_0 + \delta_1, \quad \textup{for} \, t \geq t_1
\end{equation*}

\begin{equation*}
 \| \bse_s \| \leq \rho_s + \delta_2, \quad \textup{for} \, t \geq t_2
\end{equation*}

\begin{equation*}
 \|\btau_r\| \leq a_3 \|\bss\| + a_2 \|\bsq_{e}\|^2 + a_1 \|\bsq_{e}\| + a_0 + \delta_3, \quad \textup{for} \, t \geq t_3
\end{equation*}

\begin{equation*}
\begin{array}{r}
  \|H \bsu\| \leq \rho_E b_3 \|\bss\| + \rho_E b_2 \|\bsq_{e}\|^2 + \rho_E b_1 \|\bsq_{e}\| + \rho_E b_0 + \delta_4, \\
 \textup{for} \, t \geq t_4
\end{array}
\end{equation*}
Letting $\delta = \max\{ \delta_i  \}_{i=1,2,3,4} $ and $ t_{\delta} = \max\{ t_i \}_{i=1,2,3,4} $, the result in~\ref{lem2} then follows.

%

\ifCLASSOPTIONcaptionsoff
  \newpage
\fi
\bibliographystyle{IEEEtran}
\bibliography{ref_gui}

\end{document}

%% file: BoldSymbol.tex
\newcommand{\bcd}{\mbox{\boldmath $D$\unboldmath}}

\newcommand{\bcp}{\mbox{\boldmath $P$\unboldmath}}
\newcommand{\bcq}{\mbox{\boldmath $Q$\unboldmath}}

\newcommand{\bsb}{\mbox{\boldmath $b$\unboldmath}}

\newcommand{\bsd}{\mbox{\boldmath $d$\unboldmath}}
\newcommand{\bse}{\mbox{\boldmath $e$\unboldmath}}

\newcommand{\bsn}{\mbox{\boldmath $n$\unboldmath}}

\newcommand{\bsp}{\mbox{\boldmath $p$\unboldmath}}
\newcommand{\bsq}{\mbox{\boldmath $q$\unboldmath}}
\newcommand{\bsr}{\mbox{\boldmath $r$\unboldmath}}
\newcommand{\bss}{\mbox{\boldmath $s$\unboldmath}}

\newcommand{\bsu}{\mbox{\boldmath $u$\unboldmath}}

\newcommand{\betaa}{\mbox{\boldmath $\eta$\unboldmath}}

\newcommand{\btau}{\mbox{\boldmath $\tau$\unboldmath}}

\newcommand{\bpsi}{\mbox{\boldmath $\psi$\unboldmath}}
\newcommand{\bomega}{\mbox{\boldmath $\omega$\unboldmath}}